\documentclass[10pt,conference,review,anonymous]{IEEEtran}
\IEEEoverridecommandlockouts

\usepackage{cite}
\usepackage{amsmath,amssymb,amsfonts}
\usepackage{algorithmic}
\usepackage{graphicx}
\usepackage[svgnames,table]{xcolor}
\usepackage{enumitem}
\usepackage{fontawesome}
\usepackage{multirow}
\usepackage{listings}
\usepackage{mdframed}
\usepackage{framed}
\usepackage{soul}
\usepackage{relsize}
\usepackage{caption}
\usepackage{threeparttable}
\usepackage{calc} 
\usepackage{booktabs}

\soulregister{\NE}{1}
\soulregister{\PN}{1}
\soulregister{\PA}{1}
\soulregister{\texttt}{1}

\usepackage{tikzsymbols}
\usepackage{textcomp}
  \definecolor{diffstart}{named}{gray}
  \definecolor{diffincl}{HTML}{00a67d} 
  \definecolor{diffrem}{named}{orange}

\definecolor{mygreen}{rgb}{0,0.6,0}
\definecolor{mygray}{rgb}{0.5,0.5,0.5}
\definecolor{mymauve}{rgb}{0.58,0,0.82}
  
\lstdefinestyle{gitstyle}{
  language=bash,
  keywords={},
  basicstyle=\ttfamily\small,
  frame=single,
  rulecolor=\color{gray!80},
  breaklines=true,
  postbreak=\mbox{\textcolor{red}{$\hookrightarrow$}\space},
  showstringspaces=false,
  keywordstyle=\color{blue!75}\bfseries,
  commentstyle=\color{gray},
  morekeywords={git, clone, add, cherry, pick, fetch, remote, pull, push, checkout}
  captionpos=t
}

 \lstdefinelanguage{diff}{
    morecomment=[f][\color{diffstart}]{@@},
    morecomment=[f][\color{diffincl}]{+\ },
    morecomment=[f][\color{diffrem}]{-\ },
    morecomment=[f][\color{diffrem}]{<<< },
    morecomment=[f][\color{diffrem}]{=== },
    morecomment=[f][\color{diffrem}]{>>> },
    captionpos=t
}

\usepackage{hyperref}
\hypersetup{
    colorlinks=true,
    citecolor=blue,
    filecolor=black,
    linkcolor=black,
    urlcolor=gray
}

\newlength\WIDTHOFBAR
\newcommand{\nd}{\vspace{1mm}\noindent}

\newcommand*{\tool}{\texttt{RePatch}\xspace}

\newcommand*{\PA}{\texttt{PA}\xspace} 
\newcommand*{\PN}{\texttt{PN}\xspace} 
\newcommand*{\NE}{\texttt{NE}\xspace} 

\renewcommand{\lstlistingname}{\bfseries Listing}
\makeatletter
\def\fnum@lstlisting{%
  \lstlistingname
  \ifx\lst@@caption\@empty\else~\thelstlisting\normalfont\fi}%
\makeatother

\newenvironment{custombox}{\smallskip\begin{mdframed}[linewidth=1pt,innerleftmargin=5pt, innerrightmargin=5pt, innertopmargin=5pt, innerbottommargin=5pt, nobreak=true]}{\end{mdframed}\smallskip}
  
\def\BibTeX{{\rm B\kern-.05em{\sc i\kern-.025em b}\kern-.08em
    T\kern-.1667em\lower.7ex\hbox{E}\kern-.125emX}}

\begin{document}
\title{Refactoring-Aware Patch Integration Across Structurally Divergent Java Forks\\

}

\author{\IEEEauthorblockN{Daniel Ogenrwot}
\IEEEauthorblockA{\textit{Department of Computer Science} \\
\textit{Howard R. Hughes College of Engineering}\\
\textit{University of Nevada Las Vegas}\\
Las Vegas, USA \\
ogenrwot@unlv.nevada.edu}
\and
\IEEEauthorblockN{John Businge}
\IEEEauthorblockA{\textit{Department of Computer Science} \\
\textit{Howard R. Hughes College of Engineering}\\
\textit{University of Nevada Las Vegas}\\
Las Vegas, USA \\
john.businge@unlv.edu}
}
\maketitle
\thispagestyle{plain}
\pagestyle{plain}

\begin{abstract}

While most forks on platforms like GitHub are short-lived and used for social collaboration, a smaller but impactful subset evolve into long-lived forks, referred to here as variants, that maintain independent development trajectories. Integrating bug-fix patches across such divergent variants poses challenges due to structural drift, including refactorings that rename, relocate, or reorganize code elements and obscure semantic correspondence.
This paper presents an empirical study of patch integration failures in 14 divergent pair of variants and introduces \tool, a refactoring-aware integration system for Java repositories. \tool extends the RefMerge framework, originally designed for symmetric merges, by supporting asymmetric patch transfer. \tool inverts refactorings in both the source and target to realign the patch context, applies the patch, and replays the transformations to preserve the intent of the variant.
In our evaluation of 478 bug-fix pull requests, Git \texttt{cherry-pick} fails in 64.4\% of cases due to structural misalignments, while \tool successfully integrates 52.8\% of the previously failing patches. These results highlight the limitations of syntax-based tools and the need for semantic reasoning in variant-aware patch propagation.

\end{abstract}


\section{Introduction}\label{sec:intro} 

Many modern software systems are derived by copying and adapting existing codebases, a practice known as forking~\cite{Jacob:FSE:2020, businge:2018icsme, businge:emse:2021, businge:saner:2022, pareco:2022, businge:blockchain:2022, Stefan:2016:icsme, Fenske:SANER:2017, zhou2020collaboration}. Although many of such forks are short-lived or used for temporary collaboration, a smaller but impactful subset develops into long-lived forks, referred to in this work as \textit{variants}~\cite{businge:2018icsme, businge:emse:2021, businge:saner:2022, pareco:2022, Businge:benevol:2020,Businge-book-2023,njima2020empirical}, that maintain independent development trajectories.
These variants enable customization and organizational control~\cite{businge:emse:2021, pareco:2022,lillack:2019}, but the structural and semantic divergence over time complicates the reuse of upstream patches such as bug fixes or enhancements. 

Structural and semantic divergence poses a major barrier to patch propagation. As variants undergo refactoring, such as method renaming, class movement, or interface modification~\cite{Dig:MolhadoRef:OOPSLA:2006, refmerge:2023, pan:ISSTA:2024}, syntactic similarity with the source deteriorates. Tools like Git \texttt{cherry-pick}, which rely on textual alignment, often fail when entities have been reorganized or renamed~\cite{refmerge:2023, pan:ISSTA:2024}. Even minor structural edits can cause merge conflicts or incorrect applications, requiring developers to manually identify equivalent code regions, which is a tedious and error-prone task~\cite{pareco:2022,variantsync:2016}. Although LLM-based techniques like Pan et al.~\cite{pan:ISSTA:2024} offer a promising direction for patch adaptation, they provide limited transparency in how structural edits are resolved. 
Likewise, ML-based merge tools such as \texttt{DeepMerge}~\cite{deepmerge:2020} and \texttt{MergeBERT}~\cite{mergebert:2022} frame merge resolution as token classification or sequence generation tasks over textual representations of code. Although they achieve strong results on benchmark datasets, their core models do not explicitly reason about structural drift or semantic refactorings, instead relying on learned token-level patterns. In addition, these models offer limited interpretability as their predictions lack insight into how structural changes influence merge outcomes.

Empirical studies highlight the practical challenges of patch integration. Businge et al.~\cite{businge:emse:2021} analyzed over 9{,}000 \textit{software families}\footnote{A software family refers to a group of variants that originate from a common codebase and evolve independently over time.} and found that integration between forked variants was rare, revealing the lack of effective patch propagation mechanisms. Ramkisoen et al.~\cite{pareco:2022} reported over 1{,}000 cases each of ``missed opportunities'' (patches present in the source but missing in the target variant), and ``effort duplication'' (manual re-implementation) across 364 variant pairs. Many missed patches lagged behind by a median of 52 weeks, showing the fragility of patch reuse in clone-and-own ecosystems. 
Refactoring-aware tools like \texttt{MolhadoRef}~\cite{Dig:MolhadoRef:OOPSLA:2006}, \texttt{RefMerge}~\cite{refmerge:2023}, and \texttt{IntelliMerge}~\cite{intellimerge:2019} have reduced merge conflicts in symmetric divergence scenarios where branches share recent history. However, they are not designed for asymmetric divergence, where forks evolve independently and accumulate unsynchronized structural changes. Therefore, these tools do not address the challenge of patch transfer across divergent repositories.

To address these limitations, we investigate patch integration across structurally divergent software variants. These variants share a common origin, but have evolved independently, introducing structural changes that hinder patch reuse. Rather than merging entire branches, we focus on applying individual bug-fix commits previously identified as relevant but not integrated into target variants~\cite{pareco:2022}. We examine how method-level refactorings affect patch transferability and present a refactoring-aware strategy that enhances Git’s \texttt{cherry-pick} with semantic reasoning. Our approach builds on the invert-and-replay strategy of Ellis et al.~\cite{refmerge:2023}, extending it to support integration between long-diverged repositories. 
Using \texttt{RefactoringMiner}~\cite{tsantalis2018accurate}, we detect and invert refactorings in both source and target variants to enable patch application, then replay them to preserve structural intent. This supports a systematic assessment of how refactorings affect patch integration success across real-world divergent variants.

We evaluate our approach on 478 bug-fixing patches from the \texttt{PaReco} dataset~\cite{pareco:2022}, which catalogs missed integration opportunities across divergent variant pairs. Each case uses the merge commit from the original pull request as the integration unit. Our evaluation spans 14 variant pairs. While Git \texttt{cherry-pick} fails on 64.6\% of these due to structural conflicts, our method successfully integrates 52.8\% of the previously failing patches, substantially reducing manual effort. By explicitly modeling the structural changes that cause cherry-pick failures, our approach enables interpretable and reliable patch propagation across variant-rich ecosystems. To our knowledge, this is the first empirical study to isolate refactorings as a root cause of integration failure and assess their effect on patch reuse.

\noindent \textbf{Contributions.} This paper makes the following contributions:

\begin{itemize}[leftmargin=*]
    \item We frame patch integration between long-diverged forks as a refactoring-aware cherry-pick problem and analyze the structural barriers to reuse across independent repositories.
    
    \item We present \tool, a rule-based integration system that extends \texttt{RefMerge} to asymmetric settings by performing dual-sided refactoring detection, inversion, semantic patch application, and transformation replay.

    \item Our study of 478 bug-fix pull requests across 14 Java variant pairs reveals that 91.6\% of Git cherry-pick failures are due to target-side refactorings. \tool resolves 52.8\% of previously failed cases and reduces conflicts in over half of them, while introducing new conflicts in less than 4\%.

    \item We release all tools, datasets, patch traces, and integration logs to support replication and future work on variant-aware patch propagation~\cite{scam:2025:replication}.
\end{itemize}

\begin{figure*}
    \centering
    \includegraphics[width=1\linewidth]{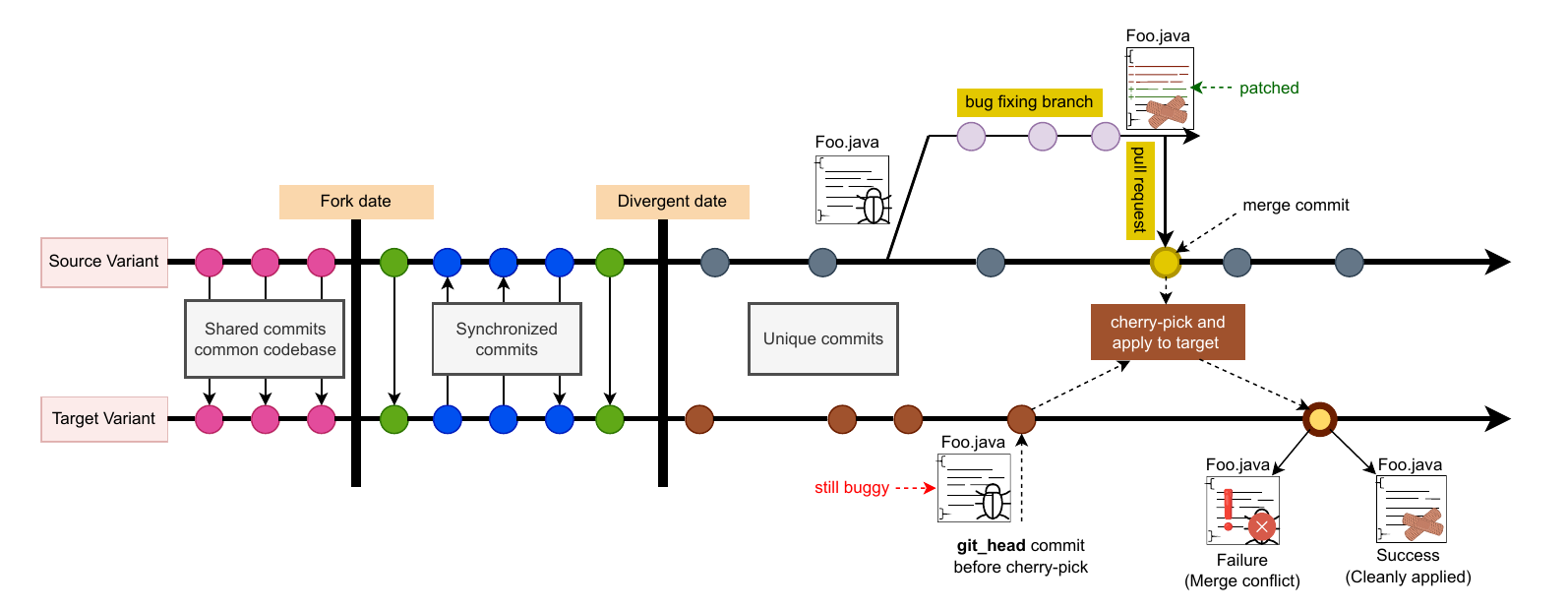}
    \caption{Illustration of patch integration from source to target variant.}
    \label{fig:illustration}
\end{figure*}

\section{Terminology, Problem and Concrete Examples} \label{sec:termi}

\nd In this section, we formally define the terminology used throughout our study and illustrate the practical challenges of patch integration in divergent forks through a concrete example.

\subsection{Terminology for Variant-Based Patch Integration} \label{sec:2-B}

\begin{itemize}[leftmargin=*]

    \item \textbf{source variant.} The repository where a patch originates, typically where the bug fix was authored and merged.
    \item \textbf{target variant.} The repository receiving the patch. Structural divergence may prevent direct application.
    \item \textbf{hunk.} A contiguous block of line-level changes in a diff, marked by \texttt{@@ -s,n +s,n @@}~\cite{hunks, patchtrack:ase:2024, patchtrack:arxiv:2025}.
    \item \textbf{patch.} A set of hunks spanning one or more files, representing all code changes in a pull request for a bug fix.
    \item \textbf{refactoring.} A behavior-preserving structural transformation (e.g., Rename Method, Move Class) that may obscure traceability between variants.

    \item \textbf{symmetric divergence.} Two branches evolve from a recent common ancestor with aligned histories, allowing bidirectional reasoning using standard merge tools~\cite{refmerge:2023}.

    \item \textbf{asymmetric divergence.} Variants evolve independently from a shared base, accumulating unsynchronized commits. Lacks a recent common ancestor, requiring semantic reasoning for patch transfer.

 \end{itemize}   

\subsection{Motivating Problem Scenario}
\label{Sec1.1:Illustration}
\noindent Figure~\ref{fig:illustration} shows patch integration between two variants that initially share a synchronized history but diverge at a specific point (\texttt{divergence\_date}). After this point, both variants evolve independently and no longer share synchronized commits.
Suppose a developer identifies and fixes a bug in \texttt{Foo.java}, a file present in both source and target variants, by submitting a pull request in the source repository. Although the same bug likely exists in the target variant, there is no direct mechanism to transfer the patch. Developers often use Git’s \texttt{cherry-pick} to apply the source commit to the target’s \texttt{git\_head}, with two possible outcomes:

\begin{enumerate}
    \item[(i)] The patch applies cleanly if the change involves simple textual edits.
    \item[(ii)] The patch fails if structural changes in the target, such as refactorings or reorganizations, conflict with the assumptions of the patch.
\end{enumerate}

\noindent Even small independent edits can cause patch failures due to mismatches in method signatures, class structures, or identifier names. Success often depends on the structural consistency between the merge commit in the source and the current state of the target. When alignment exists, patch transfer may succeed with minimal effort. Otherwise, developers must manually resolve discrepancies. Addressing this requires detecting and aligning structural edits such as refactorings. The integration in Figure~\ref{fig:illustration} is bidirectional; either variant may produce reusable patches.

\subsection{Patch Integration Examples: Success and Failure}
\noindent To illustrate these challenges, we present examples of successful and failed patch integrations from \texttt{apache/kafka} to its industrial variant \texttt{linkedin/kafka}. We begin with a success case that demonstrates how structural consistency enables Git \texttt{cherry-pick} to apply a patch cleanly. This is followed by a failure case where divergence obstructs integration.

\noindent \textbf{Example 1: Success Case.} We begin with a scenario where a source patch integrates cleanly into the target variant using Git’s native \texttt{cherry-pick}, without requiring semantic reasoning. In pull request \href{https://github.com/apache/kafka/pull/13032}{KAFKA-14540}~\cite{pr-13032}, the source repository fixes a bug in the \texttt{writeByteBuffer} method of the \texttt{DataOutputStreamWritable} class. The bug involved incorrectly passing \texttt{buf.limit()} instead of \texttt{buf.remaining()} as the length argument in the \texttt{out.write(...)} call, which could lead to data corruption when writing \texttt{ByteBuffer} content to an output stream.
\vspace{-0.3em}
\begin{lstlisting}[xleftmargin=3em, framexleftmargin=1.9em, language=diff,
caption={Diff file for PR--13032 from the source repository}, label={listing: diff_example_success},
frame=lines, basicstyle=\ttfamily\scriptsize, breaklines=true, numbers=left, stepnumber=1]
@@ -99,7 +99,7 @@ public void writeUnsignedVarint(int i) {
    public void writeByteBuffer(ByteBuffer buf) {
        try {
            if (buf.hasArray()) {
-                out.write(buf.array(), buf.position(), buf.limit());
+                out.write(buf.array(), buf.arrayOffset() + buf.position(), buf.remaining());
            } else {
                byte[] bytes = Utils.toArray(buf);
                out.write(bytes);
\end{lstlisting}

\noindent The fix updated the call to correctly compute the offset and remaining length using \texttt{buf.arrayOffset() + buf.position()} and \texttt{buf.remaining()}, and was applied cleanly to the target variant. This successful cherry-pick was enabled by the structural consistency of the file and method across both source (Listing~\ref{listing: diff_example_success}) and the \texttt{git\_head} of the target variant (file: DataOutputStreamWritable.java, lines 99--110)\footnote{\href{https://github.com/linkedin/kafka/blob/3.0-li/clients/src/main/java/org/apache/kafka/common/protocol/DataOutputStreamWritable.java}{File: \texttt{DataOutputStreamWritable.java}, lines 99--110}}. The patch was confined to a single method and introduced no conflicting edits, allowing Git to apply it without developer intervention (Listing~\ref{listing: diff_example_success}, line 6). This example shows that when source and target variants remain structurally aligned, \texttt{cherry-pick} can be a viable patch integration strategy. The next example illustrates a failure case, where semantically meaningful but structurally disruptive edits prevent clean application.

\nd \textbf{Example 2: Failure case.} 

\noindent Pull request~\href{https://github.com/apache/kafka/pull/12363}{\#12363}~\cite{pr-12363} in \texttt{apache/kafka} addresses a parameter misordering bug and improves metric naming and log reporting.\footnote{Patch details: \url{https://github.com/apache/kafka/pull/12363/files}} Although the changes are functionally straightforward, Git's \texttt{cherry-pick} fails to apply the patch cleanly to the \texttt{linkedin/kafka} variant due to structural divergence. The patch fails with 12 merge conflicts across 79 lines in five files: \texttt{KafkaStreams.java}, \texttt{PartitionGroup.java}, \texttt{PartitionGroupTest.java}, \texttt{StreamThread.java}, and \texttt{StreamTask.java}.\footnote{These represent content-level conflicts, not file-level.} For example, a semantic renaming of the metric field \texttt{totalBytesSensor} to \texttt{totalInputBufferBytesSensor} in \texttt{PartitionGroup.java} and \texttt{StreamTask.java} causes Git to fail to match equivalent logic. Additionally, the patch performs an inlining refactoring in \texttt{StreamTask.java}, replacing the separately declared variable \texttt{enforcedProcessingSensor} with a direct call to \texttt{TaskMetrics.enforcedProcessingSensor(...)} within the constructor's argument list (see Line 186 in the patch\footnotemark[2]). The modifications are shown below in Listing~\ref{listing: concrete_exampe_failure}.
\begin{lstlisting}[language=java, xleftmargin=2em, framexleftmargin=2em, frame=lines, breaklines=true, caption={Refactoring mismatch in \texttt{StreamTask.java}}, label={listing: concrete_exampe_failure}, basicstyle=\ttfamily\scriptsize, escapechar=!]
// source:
partitionGroup = new PartitionGroup(..., !\textcolor{red!70}{TaskMetrics.enforcedProcessingSensor(...)}!, ...);
// target:
final Sensor enforcedProcessingSensor;
enforcedProcessingSensor = TaskMetrics.enforcedProcessingSensor(...);
partitionGroup = new PartitionGroup(..., !\textcolor{red!70}{enforcedProcessingSensor}!, ...);
\end{lstlisting}

This example highlights how semantically meaningful edits (such as renaming and inlining) can break textual alignment and lead to patch integration failure across variants. A manual reconstruction suggests that a refactoring-aware tool could reconcile such conflicts in at least 4 of 5 affected files (80\%), reducing conflicting lines from 79 to approximately 20. However, manual resolution does not scale for large or complex patches, which underscores the need for automated support.
\section{Study Design}\label{sec:studydesign}
\noindent This section presents the research method from data collection, tool design, and experiment setup. 

\begin{figure}[ht]
    \centering
    \includegraphics[width=1\linewidth]{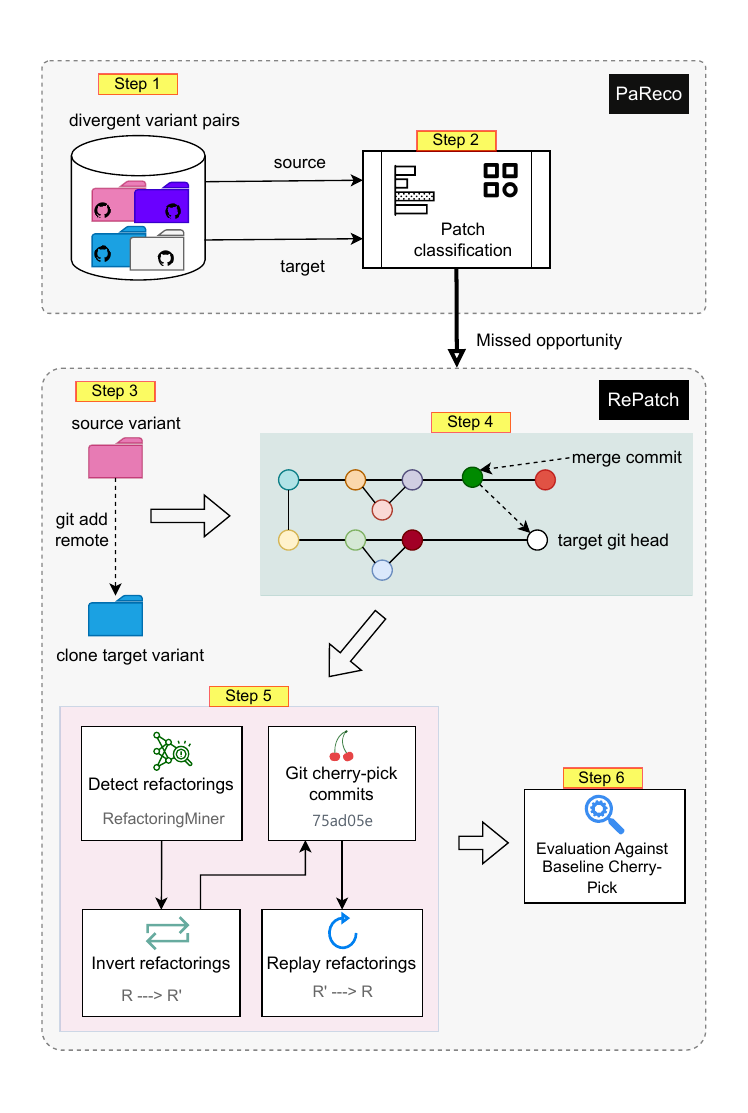}
    \caption{Overview of the research method pipeline. }
    \label{fig:method}
\end{figure}

\subsection{Research Questions}
\begin{itemize}[leftmargin=*]
    \item \textbf{RQ1:} \textit{How often do source variant bug-fix patches fail to apply cleanly to target variants using Git’s cherry-pick?}       
    This question establishes a baseline by quantifying the limitations of Git's patch integration in the presence of structural divergence between variants. It highlights the practical challenges of cross-variant patch transfer and motivates the need for refactoring-aware support.

  \item \textbf{RQ2:} \textit{What proportion of cherry-pick failures are attributable to refactoring operations (e.g., method / class renaming or moving)?}  
This question quantifies how often semantic changes, such as method or class renaming and movement, are responsible for patch integration failures. It distinguishes refactoring-related failures from those caused by deletions, interface drift, or unrelated edits, and motivates the need for refactoring-aware integration techniques.

\item \textbf{RQ3:} \textit{Can a refactoring-aware integration approach reduce merge conflicts and increase the success rate of applying patches across divergent variants?}  
This question assesses whether our method outperforms Git’s \texttt{cherry-pick} by reducing conflicts and increasing success rates. We investigate if incorporating refactoring awareness enables better integration outcomes, regardless of whether the patch originates from the source or target variant.

\end{itemize}

\subsection{Data Collection}
\noindent This section describes the steps we followed to collect the data used in our study. A visual summary of the process is presented in Figure~\ref{fig:method}.

\nd \textbf{Step 1: Selection of divergent variant pairs:}
We base our study on the dataset created by Ramkisoen et al.~\cite{pareco:2022}, which includes 364 GitHub variant pairs identified as structurally divergent. In the original study, projects were considered active if they had commit activity within the six months leading up to August 6, 2021, which marked the end of their data collection period. In our study, we reuse this dataset but apply a new activity filter to ensure that the projects are still relevant. Specifically, we retain only the variant pairs that show ongoing development activity on May 15, 2025. From this filtered set, we further restrict our selection to Java-based pairs since our refactoring-aware integration approach is implemented for Java projects. This results in a curated set of active and structurally divergent Java variant pairs for evaluation.

Table~\ref{tab:data} presents the metadata of the 14 divergent variant pairs selected for this study. Each pair consists of a source repository and a structurally divergent target variant. For each pair, we report the number bug fixing pull requests, \textit{Missed Opportunity} 
patches identified using the \texttt{PaReco} tool, as well as the number of commits the target is ahead and behind relative to the source. We also include popularity metrics (stargazers and forks) for each source repository to highlight their relevance and community engagement. The \textit{Divergent Date} indicates the estimated point at which the two repositories began to evolve independently, calculated based on the last shared commit in their histories. Since patch integration can be attempted in both directions (i.e., source$\rightarrow$target and vice versa), we treat each directional pair independently, resulting in a total of 14 divergent variant pairs for analysis. This subset provides realistic cases of structural divergence and refactoring evolution suitable for evaluating semantic patch integration across forks. 

\noindent \textbf{Step 2: Patch Identification.} 
We use the \texttt{PaReco} tool~\cite{pareco:2022} to identify bug-fix patches suitable for integration across structurally divergent Java-based variant pairs. PaReco is a token-based clone detection tool extended to analyze both modified and unmodified code regions between divergent variants. Specifically, it focuses on pull requests that contain bug-fix commits and classifies each patch based on whether its changes have been fully, partially, or not at all propagated to the target variant. This classification is determined by examining whether the modified code fragments, known as \textit{hunks}, from the source are present in the corresponding files in the target. Among these, we focus exclusively on \textit{Missed Opportunities}, defined as patches where at least one modified hunk is absent from the target variant. This includes both entirely unpropagated patches and partially applied ones that still exhibit semantic divergence. These cases offer a realistic and challenging context for evaluating the effectiveness of refactoring-aware integration techniques.

To ensure that selected patches reflect current and actionable integration scenarios, we rerun \texttt{PaReco} on each divergent variant pair using commits made from the established divergence point through May 15, 2025. For each missed opportunity identified, we extract the corresponding \texttt{merge commit} from the source repository bugfix pull request as the integration unit. The \texttt{PaReco} classification has been empirically validated with a precision of 91\%, recall of 80\%, accuracy of 88\%, and an F1 score of 85\%, confirming its reliability in identifying integration gaps where structural differences and refactorings can prevent direct reuse of upstream patches.

\begin{table*}
\renewcommand{\arraystretch}{1.25}
\centering
\begin{threeparttable}
\caption{Metadata of source and target variant Pairs used in this study.}
\label{tab:data}
\begin{tabular}{lrrlrrrrr}
\toprule
\textbf{Source} & \textbf{Stars} & \textbf{Forks} & \textbf{Target} & \textbf{Ahead\tnote{1}} & \textbf{Behind\tnote{2}} & \textbf{Diverged On} & \textbf{PRs} & \textbf{MOs} \\
\midrule
 apache/kafka~\cite{apache-kafka} & 30.1k & 14.4k & linkedin/kafka & 470 & 6,332 & 2022-06-02 & 1,247 & 393 \\
 DSpace/DSpace~\cite{dspace} & 968 & 1.3k & ufal/clarin-dspace & 1,902 & 13,418 & 2020-08-07 & 465 & 53 \\
 bitcoinj/bitcoinj~\cite{bitcoinj} & 5.1k & 2.5k & bisq-network/bitcoinj & 229 & 2,126 & 2019-03-01 & 13 & 6 \\
 bitcoinj/bitcoinj~\cite{bitcoinj} & 5.1k & 2.5k & langerhans/dogecoinj-new & 143 & 1,841 & 2019-03-01 & 13 & 6 \\
 javacc/javacc~\cite{javacc} & 1.2k & 251 & tulipcc/ParserGeneratorCC & 414 & 592 & 2018-11-20 & 40 & 2 \\
 xerial/sqlite-jdbc~\cite{xerial} & 3.0k & 641 & Willena/sqlite-jdbc-crypt & 362 & 16 & 2020-12-30 & 35 & 1 \\
 typetools/checker-framework~\cite{typetools} & 373 & 1.1k & eisop/checker-framework & 1,634 & 2,566 & 2024-09-05 & 40 & 5 \\
\midrule
\multicolumn{9}{c}{ \textbf{Source $\rightarrow$ Target Interchanged}}   \\ 
\multicolumn{9}{c}{ }   \\
 linkedin/kafka~\cite{linkedin-kafka} & 147 & 58 & apache/kafka & 6,332 & 470 & 2022-06-02 & 118 & 7 \\
 ufal/clarin-dspace~\cite{clarin-dspace} & 28 & 19 & DSpace/DSpace & 13,418 & 1,902 & 2020-08-07 & 12 & 1 \\
bisq-network/bitcoinj~\cite{bisq-network} & 14 & 28 & bitcoinj/bitcoinj & 2,126 & 229 & 2019-03-01 & 2 & 0 \\
langerhans/dogecoinj-new~\cite{dogecoinj-new} & 58 & 41 & bitcoinj/bitcoinj & 1,841 & 143 & 2019-03-01 & 0 & 0 \\
 tulipcc/ParserGeneratorCC~\cite{ParserGeneratorCC} & 10 & 7 & javacc/javacc & 592 & 414 & 2018-11-20 & 1 & 0 \\
 Willena/sqlite-jdbc-crypt~\cite{sqlite-jdbc-crypt} & 190 & 36 & xerial/sqlite-jdbc & 16 & 362 & 2020-12-30 & 7 & 1 \\
 eisop/checker-framework~\cite{eisop} & 22 & 24 & typetools/checker-framework & 2,566 & 1,634 & 2024-09-05 & 40 & 3 \\ \hline
 \multicolumn{7}{r}{ \textbf{Total} }& \textbf{2,033} & \textbf{478}  \\
\bottomrule
\end{tabular}
\begin{tablenotes}
\footnotesize
\item[1] Commits present in the target repository but absent in the source.
\item[2] Commits present in the source repository but absent in the target.
\end{tablenotes}
\end{threeparttable}
\end{table*}

\subsection{Patch Integration}
\noindent We employ \tool, a refactoring-aware patch integration system that extends \texttt{RefMerge}~\cite{refmerge:2023}. While \texttt{RefMerge} operates at the branch level and performs full-tree merges, \tool supports commit-level cherry-pick workflows suited for structurally divergent variants.
\tool adapts \texttt{RefMerge}’s core mechanisms of refactoring inversion and replay for localized, commit-scoped integration. It detects and inverts refactorings on both source and target variants, generates a clean diff from the base of the source commit, and applies it to the inverted target. Refactorings from both sides are then replayed to restore the intended structure.
To ensure semantic correctness, \tool reuses \texttt{RefMerge}'s conflict matrix, a data structure that captures ordering dependencies and replay constraints, but reconfigures it for commit-level patching. Unlike \texttt{RefMerge}, which reasons over entire branch histories, \tool applies the matrix at finer granularity, allowing it to detect and resolve structural clashes that arise during cherry-pick. This design preserves patch semantics and variant integrity while extending refactoring-aware integration to real-world cross-repository workflows.

\noindent \textbf{Step 3: Link Target and Source Repository.}  
To facilitate patch transfer, \tool begins by cloning the target repository. It then configures the source repository as a remote within the target’s Git environment. This linkage is established using the \texttt{git remote add} and \texttt{git fetch} commands, which import the source's commit history into the local Git namespace. This setup allows us to reference and operate on source commits directly within the integration context, thereby enabling cherry-pick-based patch transfer between structurally divergent repositories.

\begin{lstlisting}[style=gitstyle, basicstyle=\ttfamily\scriptsize, frame=lines]
git clone https://github.com/linkedin/kafka # target
git remote add kafka https://github.com/apache/kafka # source
git fetch kafka
\end{lstlisting}

\noindent \textbf{Step 4: Apply Bug-Fix Commits via Cherry-Pick.}  
Once the repositories are linked, we identify the relevant bug-fix commits corresponding to \textit{Missed Opportunities}, previously extracted as \texttt{merge commits} from source variants' pull requests. These commits encapsulate fixes that are present in the source variant but missing in the target.
To enable fine-grained, commit-level integration, we adopt Git’s \texttt{cherry-pick} operation as the core mechanism of our pipeline. Unlike \texttt{git merge}, which combines entire branch histories and assumes shared ancestry, \texttt{cherry-pick} allows isolated commits to be transferred across divergent variants. This makes it more appropriate for integrating individual patches without relying on symmetric version histories. 
Other tools like \texttt{git apply}, which operate directly on patch files, lack support for commit metadata, and do not integrate well with Git’s conflict resolution mechanisms. 

In contrast, \texttt{cherry-pick} offers a more reproducible and semantically meaningful integration process. For each \textit{Missed Opportunity} case, we apply the corresponding merge commit to the target variant using Git's \texttt{cherry-pick --no-commit} flag to allow for conflict handling and further transformation if needed.

\begin{lstlisting}[style=gitstyle,basicstyle=\ttfamily\scriptsize, frame=lines]
git cherry-pick <commit-hash> --no-commit
\end{lstlisting}
For each \texttt{cherry-pick} attempt, we automatically record whether the operation succeeds or fails. Failures are defined by the presence of Git-reported merge conflicts that prevent the patch from being applied cleanly. To ensure reproducibility, each \texttt{cherry-pick} is performed on a fresh checkout of the target variant at the same commit immediately prior to the attempted integration (i.e., the \texttt{git\_head} of the target variant). If the \texttt{cherry-pick} completes without conflicts, the bug fix is integrated seamlessly into the target repository. However, due to structural drift often caused by independent refactorings, class reorganizations, or identifier renaming, many such attempts result in merge conflicts. These failures signal the inadequacy of traditional syntax-based patch application in variant-aware development settings and motivate the need for semantic-aware resolution strategies, as explored in the subsequent steps of our methodology.

\noindent \textbf{Step 5: Refactoring-Aware Conflict Resolution.}  
For commits that fail to apply cleanly due to merge conflicts, we employ a refactoring-aware resolution strategy. In many cases, integration failures stem from semantic discrepancies such as method renaming, relocation, or parameter reordering that introduce structural misalignment between the source and target repositories. To address these conflicts, we use \texttt{RefactoringMiner}~\cite{tsantalis2018accurate} to detect structural refactorings in both the \texttt{cherry-picked} commit and the current \texttt{git\_head} of the target repository (See Figure~\ref{fig:illustration}). The detected refactorings are classified and processed by a rule-based algorithm implemented in \tool. The system automatically determines whether a failed integration is refactoring-related by applying deterministic matching rules (e.g., identifier correspondence, method relocation mapping) without subjective human judgment. 
These refactorings are then temporarily inverted to align the two repositories structurally, approximating the state they would have shared at the point of divergence. Simple cases such as \textit{Rename Method} or \textit{Move Method} are handled by direct identifier substitution and class relocation. For example, if a method has been renamed in the target repository from \texttt{m1} to \texttt{m2} $\rightarrow$ \texttt{\{RenameMethod (m1, m2)\}}, we temporarily invert this transformation by applying \texttt{\{RenameMethod (m2, m1)\}}. This reverts the method to its original identifier, thereby aligning the naming convention with the source patch to facilitate successful integration. For more complex cases, such as \textit{Extract Method}, \tool leverages IntelliJ IDEA’s PSI-based refactoring APIs to perform semantic inlining. Specifically, the system identifies the extracted method in the PSI tree, locates all invocation sites in the original calling method, determines the surrounding statement context, and then programmatically applies IntelliJ’s \texttt{InlineMethodProcessor} to revert the extraction (more details are povided in the \texttt{RefMerge} study~\cite{refmerge:2023}). This ensures that the original code structure matches the source patch context before reattempting integration.
Once the repositories are structurally aligned, we reapply the patch using the procedure described in \textbf{Step 4}. If the patch is successfully integrated without manual intervention, we then replay the previously inverted refactorings to restore the intended evolution history of the target repository. This two-phase process enables the integration of semantic patches even in the presence of substantial structural divergence.

\noindent \textbf{Step 6: Evaluation Against Baseline Cherry-Pick.}  
To assess the effectiveness of our refactoring-aware integration strategy, we compare each \texttt{Repatch} attempt to a baseline \texttt{Git cherry-pick} performed without any structural alignment. This comparative analysis enables us to quantify (i) the proportion of cherry-pick failures attributable to refactorings, and (ii) the extent to which our method successfully recovers failed integrations via temporary inversion and replay. These insights are critical for evaluating the practical utility of semantic awareness in automated patch transfer.
\section{Results \& Discussion}\label{sec:res-disc}

\subsection{\textbf{RQ1: How often do source variant bug-fix patches fail to apply cleanly to target variants using Git’s cherry-pick?}} \label{sec:rq1}

\noindent To answer RQ1, we quantify the baseline failure rate of Git’s native \texttt{cherry-pick} operation when applied to divergent variants. As detailed in Section~\ref{sec:studydesign}, each patch for the source variant is cherry picked into the target variant using Git’s default \texttt{cherry-pick} command without any semantic augmentation. When Git reports a merge conflict, we classify the attempt as ``Failed''; otherwise, it is recorded as ``Passed.''
Of the 478 cherry-pick attempts, 309 patches failed due to merge conflicts, resulting in an overall failure rate of 64.4\%. Table~\ref{tab:failure-rate} presents the detailed success and failure counts across all target variants. Notably, the failure rates exhibit substantial variation across projects. For instance, \texttt{clarin-dspace} experienced a failure rate of 94.3\%, and \texttt{linkedin/kafka} failed to apply 59.0\% of upstream patches. In contrast, other projects such as \texttt{dogecoinj-new}, \texttt{ParserGeneratorCC}, and \texttt{sqlite-jdbc-crypt} showed no cherry-pick failures.
To summarize the distribution across all projects, the \textbf{median failure rate is 94.3\%}, indicating that for most variants, Git’s default \texttt{cherry-pick} mechanism struggles to apply source patches cleanly. This variability reflects differing levels of structural and semantic divergence. High failure rates, particularly in \texttt{clarin-dspace} and \texttt{linkedin/kafka}, suggest significant architectural drift or refactoring activity, which disrupts the assumptions of textual equivalence required by line-based patching tools. These observations highlight the fragility of syntax-based patch propagation techniques such as cherry-pick and diff-based merging, which assume structural consistency between repositories. In fork-based development where semantic and architectural drift is common, these traditional approaches often fail, underscoring the need for refactoring-aware integration strategies.

\begin{table}[h]
\renewcommand{\arraystretch}{1.25} 
\centering
\caption{Cherry-Pick Success and Failure Rates Across Target Variants}
\label{tab:failure-rate}
\begin{tabular}{l r r r}
\toprule
\textbf{Target Variant} & \textbf{MOs} & \textbf{Passed} & \textbf{Failed} \\
\midrule
linkedin/kafka & 393  & 161 (41.0\%) & 232 (59.0\%) \\
apache/kafka   & 7    & 1 (14.3\%)   & 6 (85.7\%)   \\
\midrule
ufal/clarin-dspace & 53 & 3 (5.7\%)   & 50 (94.3\%) \\
DSpace/DSpace       & 1  & 0 (0.0\%)   & 1 (100\%)   \\
\midrule
bisq-network/bitcoinj & 6 & 0 (0.0\%)   & 6 (100\%)   \\
bitcoinj/bitcoinj     & -- & --         & --          \\
\midrule
langerhans/dogecoinj-new & 6 & 0 (0.0\%)  & 6 (100\%) \\
bitcoinj/bitcoinj         & -- & --       & --          \\
\midrule
eisop/checker-framework   & 5 & 0 (0.0\%)   & 5 (100\%)   \\
typetools/checker-framework & 3 & 1 (33.3\%) & 2 (66.7\%) \\
\midrule
tulipcc/ParserGeneratorCC & 2 & 2 (100\%) & 0 (0.0\%) \\
javacc/javacc             & -- & --       & --        \\
\midrule
Willena/sqlite-jdbc-crypt & 1 & 1 (100\%) & 0 (0.0\%) \\
xerial/sqlite-jdbc        & 1 & 0 (0.0\%) & 1 (100\%) \\
\midrule
\textbf{TOTAL} & \textbf{478} & \textbf{169} & \textbf{309} \\
\bottomrule
\end{tabular}
\end{table}

\noindent\textbf{Discussion of Findings.}  
The cherry-pick failure rate observed in our study underscores the structural fragility of patch propagation in divergent software variant ecosystems. Although the overall failure rate is 64.4\%, the median failure rate across target variants is even higher at 85.7\%, with several projects including \texttt{clarin-dspace}, \texttt{checker-framework}, and \texttt{bitcoinj} variants exhibiting failure rates between 94\% and 100\%. These results signal a fundamental barrier to automated patch reuse when repositories evolve independently through uncoordinated changes such as refactorings, code reorganizations, and identifier renamings. Even when the functional intent of a patch remains valid, Git’s syntax-driven \texttt{cherry-pick} often fails to map semantically equivalent constructs, leading to spurious conflicts or rejection of the patch.

\noindent This contrasts with symmetric development settings. For example, the \texttt{RefMerge} study~\cite{refmerge:2023} reports a median merge conflict rate of 16\% across 20 open-source projects using Git \texttt{merge}. While \texttt{merge} and \texttt{cherry-pick} serve different roles, the comparison remains informative. Git \texttt{merges} typically operate between branches with a recent common ancestor and a largely aligned structure, allowing line-based heuristics to function with relatively low conflict rates.
In contrast, cherry-pick operations across asymmetrically diverged variants must contend with semantic and architectural drift, including refactorings, identifier renamings, and code reorganizations. Even when a patch’s intent remains valid, structural misalignments can prevent Git from applying it cleanly. These findings highlight the limitations of syntax-based tools in variant-to-variant patch propagation and motivate the need for refactoring-aware integration strategies that reason beyond line-level edits. In the following sections (RQ2 and RQ3), we investigate the root causes of patch failures and introduce \tool, a system designed to improve patch transfer accuracy by reconciling structural divergence and semantic intent across independently evolving variants.

\begin{custombox}
\faLightbulbO \hspace{0.01in} \textbf{Summary of RQ1 Results}:  
\textit{The results of RQ1 demonstrate that traditional syntax-based patch integration mechanisms, such as Git’s \texttt{cherry-pick}, are inadequate for reliably propagating bug-fix patches across structurally diverged software variants. High failure rates across multiple projects reflect the compounding effects of semantic drift, architectural changes, and uncoordinated evolution. These findings establish a clear need for integration techniques that move beyond textual similarity and account for structural and semantic alignment. RQ2 and RQ3 build on this foundation by examining the nature of these failures and evaluating a refactoring-aware system designed to improve patch integration in divergent variant ecosystems.}
\end{custombox}

\subsection{\textbf{RQ2: What proportion of cherry-pick failures are attributable to refactoring operations (e.g., method / class renaming or moving)?}} \label{sec:rq2}

\nd To investigate the influence of refactorings on patch integration failures, we analyze all cherry-pick failures identified in RQ1 to determine whether these failures are semantically linked to structural changes in the target repository. For each failing patch, we apply \texttt{RefactoringMiner}~\cite{tsantalis2018accurate} to detect refactorings in both the pull request merge commit from the source and the \texttt{git\_head} of the target variant. While \texttt{RefactoringMiner} supports over 100 refactoring types, we focus our analysis on 17 structurally impactful refactorings that are supported by \texttt{RefMerge}~\cite{refmerge:2023}. These include: \textit{RenameMethod}, \textit{MoveMethod}, \textit{Move\&RenameMethod}, \textit{RenameClass}, \textit{MoveClass}, \textit{Move\&RenameClass}, \textit{InlineMethod}, \textit{ExtractMethod}, \textit{PullUpMethod}, \textit{PushDownMethod}, \textit{RenameField}, \textit{MoveField}, \textit{Move\&RenameField}, \textit{PullUpField}, \textit{PushDownField}, \textit{RenamePackage}, and \textit{RenameParameter}. A cherry-pick failure is labeled as \textit{refactoring-related} if the conflicting patch semantically overlaps with any refactored entity in the target codebase. This classification is based on syntactic references (e.g., method or class names) and structural transformations in the affected region.

\begin{figure}[h]
  \centering
  \includegraphics[width=\linewidth]{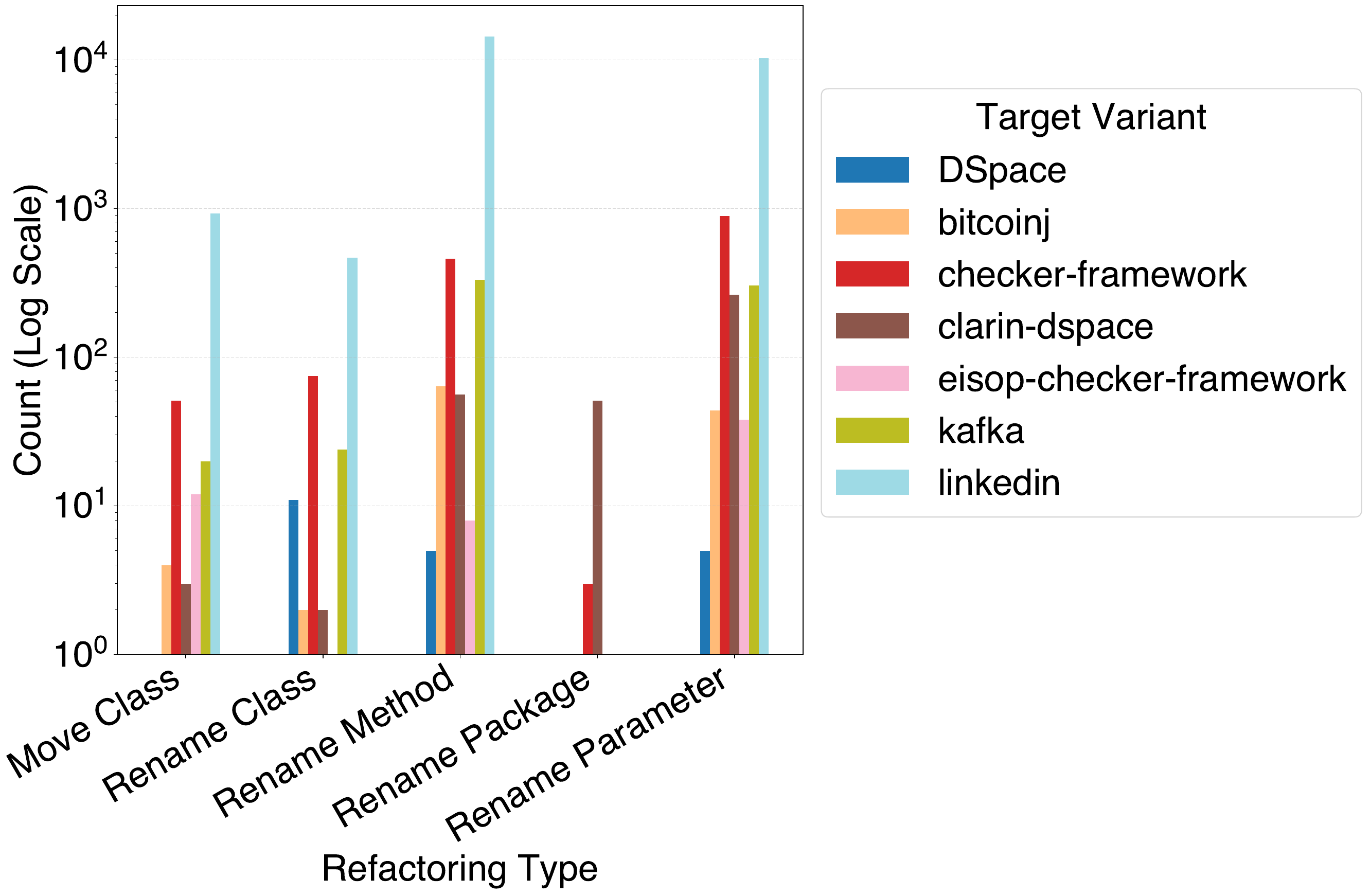}
  \caption{Top 5 refactoring types associated with cherry-pick failures, grouped by refactoring type (X-axis) and target variant (legend). Y-axis uses a log scale to normalize project size variability and improve comparability across targets.}

  \label{fig:rq2-refactor}
\end{figure}

Figure~\ref{fig:rq2-refactor} presents the five main types of refactoring that most frequently contributed to cherrypick failures between divergent variants. To ensure a meaningful interpretation, we restricted the analysis to target repositories that exhibit at least one instance of supported refactoring. The figure aggregates the number of refactoring-related conflicts per type, color-coded by project. While the figure shows the raw number of implicated refactorings, we also computed relative frequencies to identify which types are most associated with failure. The results indicate that \textit{RenameMethod (53.1\%)}, \textit{RenameParameter (41.1\%)}, and \textit{MoveClass (3.5\%)} are the most dominant contributors to conflict, especially in  \texttt{linkedin/kafka}, \texttt{apache/kafka}, \texttt{clarin-dspace}, and \texttt{typetools/checker-framework} target variants. These findings emphasize how even conceptually simple but semantically disruptive refactorings can critically undermine the applicability of patches, reinforcing the necessity for refactoring-aware strategies in modern patch management workflows.

Among the 309 cherry-pick failures identified in RQ1, 283 (\textbf{91.6\%}) were attributable to structural refactorings in the target variant. This aligns with our expectations for divergent variants, where independent evolution introduces widespread renamings, relocations, and method restructuring. In particular, projects with high refactoring activity exhibited the highest proportion of refactoring-induced conflicts. For instance, \texttt{linkedin/kafka} exhibited an 88.5\% refactoring-induced failure rate. These findings reinforce the importance of semantic traceability in maintaining long-lived forks.

\nd  
\textbf{Discussion and Implications:}  
These findings confirm that refactorings are a dominant factor behind patch integration failures in divergent forks. In most cases, the source patch references methods, classes, or identifiers that have been renamed, moved, or structurally transformed in the target, making them unresolvable through Git’s syntax-based \texttt{cherry-pick}. This limitation underscores the broader need for integration tools that incorporate semantic reasoning, especially when structural correspondence between variants has eroded over time. Our analysis shows that even seemingly simple transformations, such as method renaming or parameter reordering, can derail patch transfer when semantic alignment is not maintained.

Prior work such as \texttt{RefMerge}~\cite{refmerge:2023} demonstrated that refactoring-aware merging can significantly reduce conflicts in symmetric development scenarios, successfully resolving or reducing conflicts in 25\% of merge cases. Our study builds on this insight by demonstrating the heightened importance of refactoring-awareness in asymmetric settings, where patches from a stable source must be applied to independently evolving targets. In this context, 91.6\% of patch integration failures were attributable to structural refactorings that disrupt identifier and context alignment across source and target. This sharp contrast highlights the structural fragility of fork-based development and the limitations of syntax-driven heuristics in the presence of long-term architectural drift. These results motivate the elevation of refactoring detection and traceability as first-class capabilities in cross-variant patch integration workflows, where accurate semantic mapping is essential for bridging the gap between source intent and target structure.

\begin{custombox}
\faLightbulbO \hspace{0.01in} \textbf{Summary of RQ2 Results}:  
\textit{We find that \textbf{91.6\%} of cherry-pick failures across variant pairs are attributable to refactorings in the target repositories. The most prevalent types include \textit{RenameMethod} (53.1\%), \textit{RenameParameter} (41.1\%), and \textit{MoveClass} (3.5\%), which break identifier traceability despite semantic equivalence. Like }\texttt{RefMerge}\textit{, which reverts and replays refactorings to reduce merge conflicts in symmetric settings, our approach applies a similar strategy in an asymmetric context, where structural drift has accumulated. These results underscore the need for refactoring-aware integration tools to support reliable patch propagation across structurally divergent forks.}
\end{custombox}

\subsection{\textbf{RQ3: Can a refactoring-aware integration approach reduce merge conflicts and increase the success rate of applying patches across divergent variants?}} \label{sec:4-rq3}

\noindent To assess the effectiveness of our refactoring-aware integration, we applied \tool to all 309 cherry-pick failures identified in RQ1. \tool extends Git’s \texttt{cherry-pick} with semantic awareness via automated refactoring detection and transformation (see Section~\ref{sec:studydesign}). A patch is considered successfully integrated if it applies without textual conflicts, and resolution rate is measured as the proportion of previously failing patches that \tool integrates successfully, compared to Git’s baseline.

\tool failed to complete execution for 17 of the 309 patches due to timeouts, which we conservatively treat as failures. Manual inspection showed that these typically involved many conflicting files or extensive edits. To manage resources and ensure experimental completion, we imposed a 15-minute timeout per patch. We hypothesize that lifting this constraint could yield more successful cases. Our reported analysis thus focuses on the remaining 292 patches, excluding timeouts from resolution statistics.

\begin{table*}[t]
    \centering
    \renewcommand{\arraystretch}{1.2}
    \caption{Patch integration outcomes when applying \texttt{RePatch} to previously failing Git \texttt{cherry-pick} scenarios. The table reports the number of cases where \texttt{RePatch} reduced ($\downarrow$), increased ($\uparrow$), or unchanged ($-$) the number of conflicting files and lines of code (LOC), relative to the baseline Git cherry-pick conflict state. Only target variants with at least one cherry-pick failure are included in this analysis. 
    }

    \label{tab:outcomes}
    \begin{tabular}{l c | c c c | c c c}
        \toprule
        \multirow{2}{*}{\textbf{Target Variant}} & \multirow{2}{*}{\textbf{Total Failures}} & 
        \multicolumn{3}{c}{\textbf{Conflicting Files}} & 
        \multicolumn{3}{c}{\textbf{Conflicting LOC}} \\
        \cmidrule(lr){3-5} \cmidrule(lr){6-8}
        & & \textbf{$\downarrow$ Reduced} & \textbf{$\uparrow$ Increased} & \textbf{-- Unchanged} & 
            \textbf{$\downarrow$ Reduced} & \textbf{$\uparrow$ Increased} & \textbf{-- Unchanged} \\
        \midrule
        linkedin/kafka & 232 & 111 (47.8\%) & 4 (1.7\%) & 117 (50.4\%) 
                                 & 114 (49.1\%) & 9 (3.9\%) & 109 (46.9\%) \\
        ufal/clarin-dspace & 50 & 37 (74.0\%) & 0 (0.0\%) & 13 (26.0\%) 
                                   & 37 (74.0\%) & 0 (0.0\%) & 13 (26.0\%) \\
        apache/kafka & 3 & 3 (100\%) & 0 (0.0\%) & 0 (0.0\%) 
                              & 3 (100\%) & 0 (0.0\%) & 0 (0.0\%) \\
        typetools/checker-framework & 2 & 2 (100\%) & 0 (0.0\%) & 0 (0.0\%) 
                                             & 2 (100\%) & 0 (0.0\%) & 0 (0.0\%) \\
        bisq-network/bitcoinj & 4 & 1 (25.0\%) & 0 (0.0\%) & 3 (75.0\%) 
                                       & 0 (0.0\%) & 1 (25.0\%) & 3 (75.0\%) \\
        DSpace/DSpace & 1 & 1 (100\%) & 0 (0.0\%) & 0 (0.0\%) 
                               & 1 (100\%) & 0 (0.0\%) & 0 (0.0\%) \\
        \midrule
        \textbf{Total} & \textbf{292} & 
        \textbf{155 (53.1\%)} & \textbf{4 (1.4\%)} & \textbf{133 (45.5\%)} & 
        \textbf{157 (53.8\%)} & \textbf{10 (3.4\%)} & \textbf{125 (42.8\%)} \\
        \bottomrule
    \end{tabular}
\end{table*}

Table~\ref{tab:outcomes} shows the effectiveness of our refactoring-aware tool, \texttt{RePatch}, across 292 previously failing previously Git's \texttt{cherry-pick} patch integration scenarios. Each row reports the number of patches where Git cherry-pick failed, the number of patches where \tool reduced, increase or did not change the amount of conflicting files and lines of code (LOC) in a given target variant. For clarity, we use $\uparrow$, $\downarrow$, and $-$ to indicates instances where \tool increase, reduce or preserved the number of conflicts, respectively.

\texttt{RePatch} reduced file-level conflicts in 155 of 292 previously failing patch applications $(53.1\%)$ and line-level conflicts in 157 cases $(53.8\%)$, while increasing conflicts in only $1.4–3.4\%$ of cases. These improvements were especially notable in high-conflict variants like \texttt{linkedin/kafka} and \texttt{clarin-dspace}, where \texttt{RePatch} resolved $47.8\%$ and $74.0\%$ of failures respectively. Even in variants with fewer structural differences, RePatch consistently outperformed Git. Overall, it successfully integrated $163$ out of the original 309 failed patches, achieving a resolution rate of \textbf{$52.8\%$}.

To contextualize these results, we revisit the failure case in Example~2 (Listing~\ref{listing: concrete_exampe_failure}). In that instance, Git's \texttt{ cherry pick} produced 12 merge conflicts across 5 files, totaling 79 conflicting lines, mainly due to renaming and inlining. When processed with \texttt{RePatch}, 4 of the 5 files were resolved cleanly, and conflicts were reduced by approximately 75\% (from 79 to ~20 lines). However, \texttt{RePatch} also introduced a new conflict in \texttt{StreamThread.java}, a file that Git handled cleanly, resulting in 6 additional conflicting lines. Manual inspection attributes this to overgeneralized inlining heuristics misaligned with the target’s constructor signature.
This case mirrors broader trends in Table~\ref{tab:outcomes}, where \texttt{RePatch} reduced file-level conflicts in 53.1\% of cases (155/292) and line-level conflicts in 53.8\%, while introducing new conflicts in only 1.4–3.4\%. The example illustrates both the strengths and trade-offs of refactoring-aware integration: \texttt{RePatch} can substantially improve patch portability by reconciling semantic drift in structurally evolved forks like \texttt{linkedin/kafka}, but also highlights the need for precise transformation heuristics to avoid regressions in otherwise conflict-free files. Together with RQ2, these findings show that refactoring-induced drift is not only common (91.6\% of failures) but can be effectively mitigated through semantically informed integration.

\noindent
\textbf{Discussion and Implications:}  

Our results show that refactoring-aware integration reduces developer effort when maintaining divergent variants. \texttt{RePatch} performs well when semantic intent is preserved (e.g., via renaming or reordering), enabling low-effort patch application. These findings support temporary structural alignment and refactoring replay as practical strategies. Compared to \texttt{RefMerge}~\cite{refmerge:2023}, which resolves 25\% of merge conflicts in symmetric settings, \texttt{RePatch} targets asymmetric integration and recovers 52.8\% of failed cherry-picks. While the settings are not directly comparable, this improvement reflects \texttt{RePatch}'s tighter scope, commit-level granularity, and targeted integration workflow.

Failure cases expose limitations in semantic mapping and refactoring detection. Integration tends to fail when target-side changes delete or reorganize context-critical code, or when transformations are too subtle for current tools like \texttt{RefactoringMiner}~\cite{tsantalis2018accurate}. This reflects earlier observations on the brittleness of automated merging under structural divergency~\cite{lippe:1992, apel2011semistructured}. Although ML-based tools such as \texttt{DeepMerge}~\cite{deepmerge:2020}, \texttt{MergeBERT}~\cite{mergebert:2022}, and \texttt{PPAtHF}.~\cite{pan:ISSTA:2024} show promise, they often lack transparency. In contrast, \texttt{RePatch} offers an interpretable rule-based pipeline that supports traceability and debugging. 
Our approach aligns with previous work on semantic program transformation~\cite{meng2012lase} and reinforces the case for treating refactoring tracking as a first-class concern in integration workflows.

\begin{custombox}
\faLightbulbO \hspace{0.01in} \textbf{Summary of RQ3 Results}:  
\textit{Out of 292 previously failing patch applications, \texttt{RePatch} reduced file-level merge conflicts in \textbf{53.1\%} of cases and line-level conflicts in \textbf{53.8\%}. In contrast, it introduced new conflicts in only \textbf{1.4\%} and \textbf{3.4\%}, respectively. Overall, \texttt{RePatch} successfully integrated \textbf{155} of the \textbf{292} failed patches, achieving a resolution rate of \textbf{52.8\%}. 
These findings show that refactoring-aware integration can substantially improve patch applicability while reducing manual effort and semantic misalignment associated with structural divergence.
}

\end{custombox}

\section{Related Work}

\textbf{Fork Variants and Patch Reuse.} Forking is common in open-source and industrial development to support experimentation, customization, or business-specific needs~\cite{businge:2018icsme}. While some forks are short-lived, long-lived variants often evolve independently with little integration of upstream changes~\cite{krueger2002software, brindescu:emse:2020}. Ramkisoen et al.~\cite{pareco:2022} used the \texttt{PaReco} tool to identify two major challenges: \textit{missed opportunities}, where patches applied in one variant are absent or only partially reused in another, and \textit{effort duplication}, where similar changes are reimplemented manually. However, prior work did not provide actionable mechanisms for patch transfer, especially when structural misalignments hinder reuse. Our work fills this gap by enabling systematic recovery of unpropagated patches through a refactoring-aware integration strategy that addresses variant-specific structural drift.

\textbf{Learning-Based Patch Porting.} Traditional tools like Git's \texttt{cherry-pick} assume syntactic similarity and often fail when code structure diverges across forks. Early machine learning systems such as \texttt{LASE}~\cite{meng2012systematic, meng2012lase} learn recurring edit patterns to apply patches across similar contexts but struggle when structural alignment is lost. Deep learning-based merge tools like \texttt{PatchNet}~\cite{PatchNet:2021}, \texttt{DeepMerge}~\cite{deepmerge:2020}, and \texttt{MergeBERT}~\cite{mergebert:2022} frame conflict resolution as token-level prediction or sequence modeling, achieving high accuracy on benchmarks. However, they typically assume symmetric divergence and lack interpretability, limiting their reliability in practice. Recent LLM-based approaches like \texttt{PPatHF}~\cite{pan:ISSTA:2024} use large models for zero-shot patch porting. While they reduce reliance on historical data, their performance degrades under significant refactoring and offers little transparency into how transformations are applied. This reflects the broader challenge identified by Barr et al.~\cite{barr2012plastic}, who showed that code is often reused under renamed or structurally altered forms, making syntactic alignment unreliable.

\textbf{Refactoring-Aware Merge and Detection Tools.} Early work on operation-based merging~\cite{mens2004analyzing, lippe:1992} emphasized modeling changes as semantic operations rather than textual diffs. Building on this vision, several rule-based systems have been developed to model code structure and refactorings explicitly. \texttt{MolhadoRef}~\cite{Dig:MolhadoRef:OOPSLA:2006}, \texttt{IntelliMerge}~\cite{intellimerge:2019}, \texttt{CatchUp!}~\cite{catchup:2005}, \texttt{ReBA}~\cite{ReBA:2008}, and \texttt{RefMerge}~\cite{refmerge:2023} implement semantic merge strategies that leverage AST differencing or transformation logs to resolve conflicts more accurately than line-based tools. However, these systems assume symmetric divergence within a shared repository history and are not designed for one-sided patch integration across forks. They often rely on precise refactoring detection, such as that provided by \texttt{RefactoringMiner}~\cite{tsantalis2018accurate, Tsantalis:2022}, which identifies behavior-preserving changes like method renaming or class relocation~\cite{fowler1999refactoring, fowler2018refactoring, mens:refactoring:2004}. Empirical studies have shown that refactorings can increase the likelihood of fault-prone commits~\cite{bavota:bugs:2012, Ferreira:2018, massimiliano:bugs:2020}, yet few systems operationalize this knowledge for cross-repository patch reuse. Our approach, \texttt{RePatch}, addresses this gap by extending \texttt{RefMerge} to operate at the granularity of individual patches and support asymmetric integration through patch-scoped refactoring inversion and replay.

\textbf{Positioning Our Contribution.} While prior work has diagnosed the challenges of patch propagation across forks, including integration failures due to structural drift~\cite{pareco:2022}, few tools provide actionable mechanisms to recover and apply unpropagated patches. Our contribution lies in bridging this gap through a refactoring-aware rule-based system that enables reproducible and interpretable integration of missed patches, even across repositories with no recent shared history.

\section{Threats to Validity} \label{sec:threats}

\noindent
Our findings are subject to a number of threats to validity.

\noindent
The reliability of our semantic patch integration depends on accurate refactoring detection. We use \texttt{RefactoringMiner}~\cite{tsantalis2018accurate}, a widely adopted Java refactoring tool with high reported precision and recall. However, like all static analyzers, it may miss or misclassify fine-grained or intertwined refactorings, which can affect the correctness of inversion and replay. Our current study focuses on 14 Java-based variant pairs and does not yet include multilingual or proprietary systems. To broaden applicability, we are extending \texttt{RePatch} to support additional languages using parser-agnostic tools such as \texttt{Tree-sitter} and AST differencing.

\noindent
Third, each patch integration attempt was subject to a 15-minute timeout to ensure experimental feasibility. 17 patches (5.5\%) exceeded this limit, typically due to high file churn or extensive structural edits, and were conservatively treated as failures. Future work will investigate adaptive timeout strategies and analyze computational tradeoffs.
Fourth, in rare cases, \texttt{RePatch} introduced conflicts not detected by Git \texttt{cherry-pick}, likely due to overgeneralized heuristics or incomplete refactoring mappings. We are improving our matching logic and introducing validation checks to increase precision and reduce false positives.
Fifth, our evaluation focused on syntactic conflict resolution and did not perform functional validation of integrated patches (e.g., via test suites). Future work will assess behavioral correctness.
Finally, our classification of refactoring-induced failures is based on static analysis and transformation logs. While scalable, this approach may not fully capture developer intent in ambiguous cases. To promote transparency and reproducibility, we publicly release our dataset, scripts, and logs~\cite{scam:2025:replication}.
\section{Conclusion}\label{sec:conc}

\noindent
This study presents the first empirical analysis of patch integration failures across long-lived, structurally divergent forks, isolating refactorings as a key source of conflict. By modeling integration as a refactoring-aware process, we demonstrate that semantic alignment significantly improves patch applicability over Git’s syntactic \texttt{cherry-pick}, resolving 52.8\% of previously failing cases across 14 real-world variant pairs.

To enable this analysis, we extend \texttt{RefMerge} with commit-level, bidirectional refactoring tracking and replay, implemented in our prototype tool \tool. These findings highlight the need for semantic reasoning in variant-aware development and suggest directions for future work, including multi-language support and integration into patch management pipelines.

\bibliographystyle{IEEEtran}
\typeout{}
\bibliography{biblio}


\end{document}